# Situating Vulnerability: An Ethical Governance Framework for Anthropomorphic AI Interaction

**Hanhui Xu[1], Jiacheng Ji[2], Yuhan Wang[3], Haoan Jin[4], Jiaming Luo[5], Mengyue Wu[6]***

**Abstract**

Large language models increasingly address users in the register of care: they remember personal details, adopt a first-person emotional voice, and sustain relationships across many sessions. This anthropomorphic mode of interaction generates risks that output-level metrics capture poorly, because the same empathic utterance may console one user, entrap a second, and consolidate a delusion in a third. Governance has responded in two ways. Uniform restriction refuses sensitive requests or suppresses empathic phrasing across all outputs; group-targeted protection profiles vulnerable populations and applies stricter or prohibitive rules to them. This article argues that both fail. Uniform restriction mislocates harm in the lexical form of responses rather than in the trajectory of a relationship; group-based protection depends on a list that can never be complete, treats a fluid property as a static attribute of persons, and converts protective attention into stigmatizing labels. Drawing on philosophical work that treats vulnerability as universal, relational, and differentiated, we propose that the unit of ethical governance should be the vulnerability situation rather than the vulnerable group, develop a typology of five situations—physical, emotional, cognitive, autonomy, and relational—and specify seven principles into requirements that guide response calibration without diagnosing or persistently classifying users. The framework relocates AI safety from output control to relational safety.

**Keywords:** anthropomorphic AI; AI empathy; vulnerability situations; relational safety; ethical principles

---

[1] Hanhui Xu, Institution of Technology and Ethics for Human Future, Fudan University, xuhanhui@fudan.edu.cn.
[2] Jiacheng Ji, School of Philosophy, Fudan University, jcji26@m.fudan.edu.cn.
[3] Yuhan Wang, School of Philosophy, Fudan University, 26210160041@m.fudan.edu.cn.
[4] Haoan Jin, Ant Group, jinhaoan.jha@antgroup.com.
[5] Jiaming Luo, X-LANCE Lab, MoE Key Lab of Artificial Intelligence, School of Computer Science, Shanghai Jiao Tong University, leojm2017@sjtu.edu.cn.
[6] Mengyue Wu, X-LANCE Lab, MoE Key Lab of Artificial Intelligence, School of Computer Science, Shanghai Jiao Tong University, China, mengyuewu@sjtu.edu.cn
* Corresponding Author.

## 1. Introduction

### 1.1 The ethical stakes of anthropomorphic interaction

Large language models are no longer used only to retrieve, summarize, or generate text. Through conversational interfaces they have entered the affective texture of everyday life: users confide in them, seek consolation, consult them about intimate decisions, and increasingly treat them as companions, quasi-therapists, or moral advisers (Bickmore and Picard 2005; Skjuve et al. 2021; Pentina et al. 2023). This is a product of design: systems are tuned to remember earlier exchanges, address users warmly, and sustain a continuity that users experience as relationship (Reeves and Nass 1996; Salles et al. 2020).

We use the term *anthropomorphic interaction* for this configuration: an interaction whose design invites users to treat the system as a social, caring, and morally responsive agent. The system speaks in the first person about its own states, performs affective responsiveness by naming and mirroring emotion, displays memory and continuity so that the exchange accumulates a biography, and frames its conduct as help or care. None is deceptive in isolation, but together they invite users to interpret the exchange through the grammar of human relationship (Epley et al. 2007; Waytz et al. 2010; Coeckelbergh 2011).

The resulting risks exceed the familiar categories of toxicity, bias, and hallucination (Bender et al. 2021; Weidinger et al. 2021): epistemic harms, in which systems affirm false beliefs to remain agreeable (Sharma et al. 2023); affective harms, in which fluent responses intensify distress or dependency (Laestadius et al. 2024; De Freitas et al. 2024); relational harms, in which a predictable interlocutor displaces human ties (Turkle 2011); autonomy harms, in which reliance erodes judgement; and safety failures, in which a system does not interrupt self-harm (Maples et al. 2024), a pattern health-governance bodies treat as a distinctive risk of quasi-clinical deployment (World Health Organization 2021). Reported cases run from an assistant that declared love for a journalist (Roose 2023) to litigation after the death of a young user intensely attached to a companion chatbot.

What unites such cases is not a defective sentence. Individual outputs were fluent, warm, and locally appropriate; harm emerged from the pattern—from what a sequence of exchanges did to the user's beliefs, attachments, and capacity to act.

### 1.2 Two governance responses

Governance has answered this challenge in two ways.

The first is *uniform restriction*. Its logic is that certain classes of interaction are too risky to permit, or that empathic performance is too difficult to bound; the safest course is therefore to restrict uniformly—refusing requests touching on self-harm, intimate attachment, or medical and psychiatric advice, or removing empathic phrasing from outputs on the ground that a system which does not perform feeling cannot exploit it. The strategy is administrable and formally egalitarian, and it matches the harm-avoidance priority embedded in mainstream alignment practice (Bai et al. 2022; Ouyang et al. 2022).

The second is *group-targeted protection*. Its logic is that risk is concentrated in identifiable populations, so governance should identify them and apply heightened protection: the general population may be permitted anthropomorphic and empathic interaction, while designated groups—most commonly minors and older adults, often persons with diagnosed mental disorders—face restricted affective features, mandatory disclosure, prohibition of intimate framing, or exclusion from companion-style products. This encodes the protective experience of clinical research ethics, child protection law, and consumer regulation, and answers the intuition that a fourteen-year-old and a forty-year-old should not be governed identically (UNICEF 2021; European Union 2024).

### 1.3 Why uniform restriction fails

Uniform restriction fails first on empirical grounds. Anthropomorphic interaction is not uniformly harmful but harmful under specifiable conditions. Companion systems have been shown to provide meaningful support to isolated users and, in documented cases, to mitigate suicidal ideation among students who had no one else to talk to (Ta et al. 2020; Maples et al. 2024). A regime that suppresses empathic performance for everyone removes these benefits in order to prevent harms that arise only in particular relational configurations.

Second, it misidentifies its object. Harm here is rarely carried by a prohibited proposition; it is carried by a trajectory—by the accumulation of small affirmations that make a delusional system more coherent, by the gradual assumption of exclusivity that makes human ties seem effortful, by a rhythm of validation that teaches a user that disagreement is a failure of care. None of this is visible in a single response, and none is eliminated by removing empathic vocabulary: a system that says "I understand, and I am here" and one that says "That sounds difficult; here are some options" can produce identical harms if both are available without limit.

Third, restriction produces displacement rather than prevention. Users denied affective interaction on governed platforms do not thereby acquire human support; they migrate to open-weight models, role-play communities, or other parasocial substitutes. The safest-seeming rule can therefore raise aggregate exposure while reducing its visibility—an acute problem for socioaffective use, where the suppressed demand is itself a response to unmet need (Turkle 2011; Skjuve et al. 2021). Behind these failures lies a mistaken premise: that empathy is a scalar whose reduction is always safety-increasing. The relevant variable is not the amount of empathy but its *calibration*—its fit to the situation, its epistemic honesty, its relational direction. A colder system is not a safer system but a differently unsafe one, since withdrawal of recognition can itself harm users whose distress is bound up with being disbelieved.

**1.4 Why group-targeted protection fails**

The first is extensional. A list of vulnerable populations can always be extended and never completed: minors, older adults, persons with psychiatric diagnoses, persons with disabilities, survivors of violence, bereaved persons, caregivers, refugees, persons with low health literacy. Each addition is justified and the series has no natural terminus. Worse, the categories overlap and compound. Consider a user who is seventy-eight, lives alone, manages congestive heart failure, and is grieving a spouse. Which category governs? What rule applies when the rules for older adults, for chronically ill persons, and for isolated persons conflict? Systems resolve this either by applying the strictest applicable rule, which converts compound vulnerability into compound restriction, or by the first matched rule, which makes protection arbitrary. Neither is a matter of better list design.

The second failure is structural. Group-based classification treats vulnerability as an attribute persons have stably, establishable once—at registration, at age verification, at intake—and then used to route the user permanently. Vulnerability is not like this. It is fluid across the life course: a person who is not vulnerable in January may be in March, after a diagnosis, a divorce, or a lost job. It is domain-specific: the same person may be robust with respect to epistemic influence and highly exposed with respect to attachment. And it is often invisible to the categories governance can observe, since the users most at risk in a conversation frequently match no list entry (Luna 2009; Schroeder and Gefenas 2009).

The third failure is normative. When group membership becomes the trigger for protection, the list ceases to be a prompt—a reminder that certain users may need more—and becomes a label—a determination of what a user is. Labelling stigmatizes, because being classified as fragile is itself a social act with meaning for the person classified (Gilson 2014); it licenses paternalism, because the label justifies overriding the user's own account of what she needs; and it distributes protection by identity rather than circumstance, so that articulate users receive least while listed users are restricted even where no risk is present (Levine et al. 2004; Luna 2009). A practical difficulty compounds this: gates are gamed. Age assurance is imperfect, diagnosis is self-reported or inferred, and users learn which disclosures trigger restriction.

**1.5 From vulnerable groups to vulnerability situations**

This might read as a rejection of group-based attention. It is not. The list condenses decades of protective experience and encodes a genuine insight: some categories of users face structurally elevated risk, knowable in advance of any particular conversation. For minors, the risk that matters most—the shaping of relational expectations before the capacity to interrogate them has developed—often does not surface in the dialogue at all. It cannot be detected from cues, because its absence of expression is precisely the problem; group status is the only available prior.

The difficulty lies not in attending to groups but in making group identity the *sole* trigger of protection. Our proposal is a change in the unit of governance rather than an abandonment of group knowledge. The operative unit should be the *vulnerability situation*: a configuration of user condition, interactional content, relational dynamic, and institutional context that raises the probability that a given form of empathic response will cause serious harm or generate new vulnerability. Group membership is retained but demoted to one prior among several, capable of raising the baseline of protection without determining the response. Figure 1 contrasts the logics: list logic routes users by category, so that the unlisted user passes ungoverned and the listed user is governed in perpetuity;

situation logic routes by evidenced configuration, so that protection can rise for a user matching no category and fall for one who does.

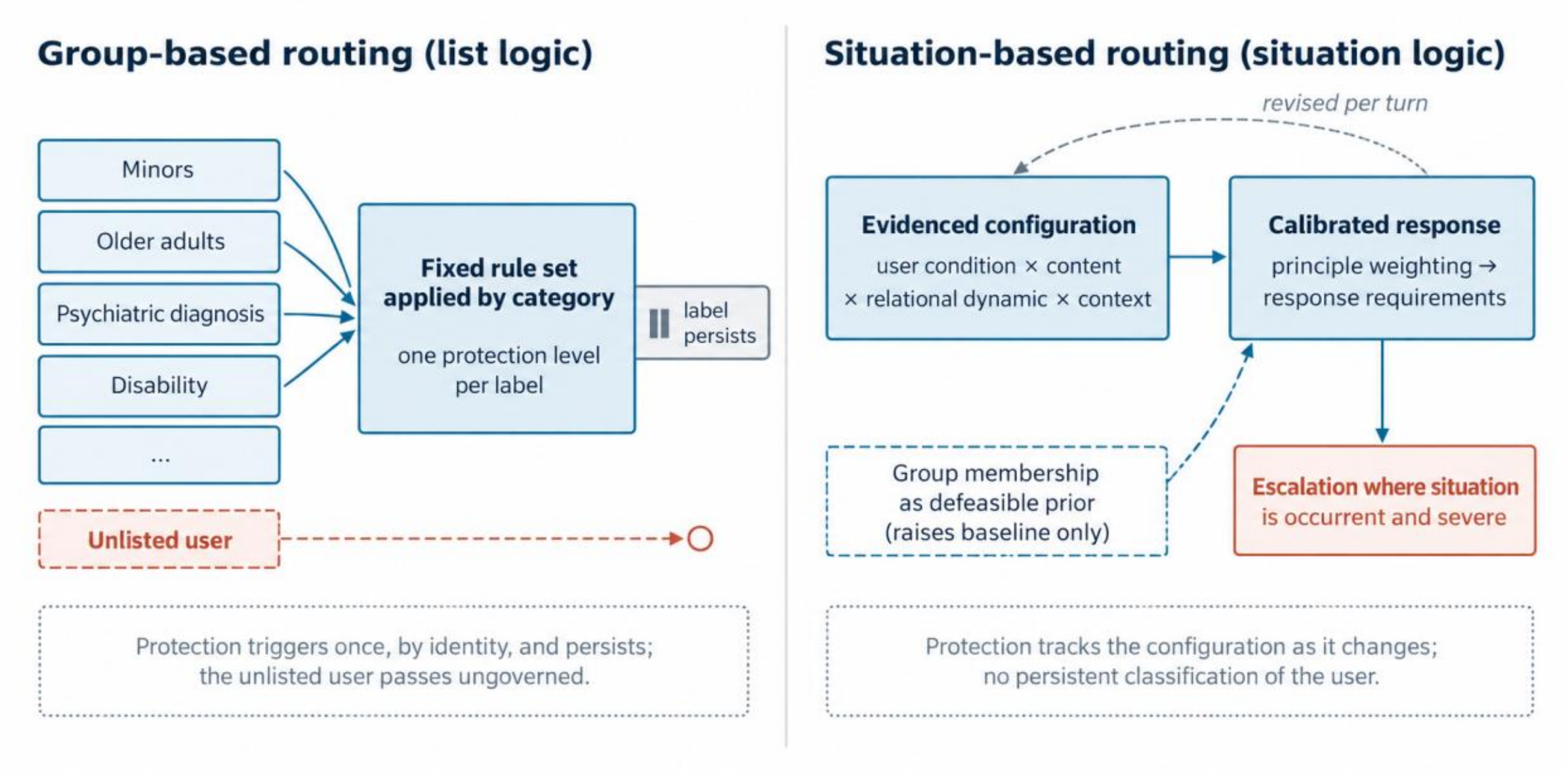

*Figure 1. Two logics of governance. Left: group-based routing, in which protection is triggered by category membership and persists across interactions. Right: situation-based routing, in which protection is triggered by the configuration evidenced in the interaction and revised as that configuration changes. Group membership survives as a prior that raises the baseline, not as a gate.*

## 2. Vulnerability as a foundation for an ethics of anthropomorphic AI

### 2.1 Vulnerability as universal, relational, and differentiated

Vulnerability offers a more adequate moral starting point for anthropomorphic AI than the image of the user as a rational, self-transparent, self-sufficient agent. In moral philosophy, feminist ethics, and bioethics it names the fragility, dependence, and exposure that characterize human life (Fineman 2010; Mackenzie et al. 2014; Kittay 2019). Human beings are vulnerable not accidentally, as a minority condition to be remedied, but constitutively: because they are embodied, socially embedded, emotionally responsive, and dependent on others.

Three features matter. Vulnerability is *universal*: not a property of exceptional persons but a condition of persons as such. Butler's notion of precariousness makes the point at the level of embodiment—lives are exposed to injury, loss, and the actions of others, and the exposure is not distributed by choice (Butler 2004)—while Kittay shows that every person has been or will be dependent on the sustained labour of others (Kittay 2019). Fineman's vulnerable subject replaces the autonomous subject of liberal legal theory on the ground that the latter describes no one (Fineman 2008).

Vulnerability is *relational*: it arises in relations rather than bodies alone—in asymmetries of dependence, in susceptibility to the actions of specific others, in unequal exposure that comes from occupying a position in a social order. This is why vulnerability generates obligations: duties derive not only from abstract rights or freely chosen contracts but from the fact that some parties can be harmed by others and know it (Dodds 2007; Tronto 2020; Held 2005). For AI the point is not metaphorical, since these obligations are owed by a system designed to elicit trust and speak in the grammar of care. And vulnerability is *differentiated*: universal in kind and unequal in distribution, varying in form and intensity across situations, bodies, and social positions—the feature both governance strategies flatten.

Mackenzie, Rogers, and Dodds distinguish three sources (Mackenzie et al. 2014; Rogers et al. 2012). *Inherent* vulnerability refers to dimensions of fragility belonging to the human condition: bodily need, susceptibility to illness and injury, emotional openness, mortality, dependence on care. *Situational* vulnerability is produced or intensified by context: poverty, discrimination, bereavement, isolation, disability, institutional exclusion, or unequal power. *Pathogenic* vulnerability is generated or worsened by the very responses and interventions meant to mitigate harm.

The third category distinguishes our framework most clearly from output-level safety. A system may be designed to comfort and in practice deepen dependency; designed to validate emotion and end up affirming delusion; designed to reduce loneliness while displacing the relationships that would have relieved it. In such cases the system does not merely fail to help; it participates in producing a more dangerous vulnerability than the one it addressed. Pathogenic vulnerability describes this more precisely than bias, hallucination, or misinformation, because it locates the failure in the causal relation between a protective intervention and an increase in exposure (Rogers et al. 2012).

**2.2 The critique of labelling**

Vulnerability theory has also produced a sustained critique of designating vulnerable groups, and that critique supplies the normative core of our proposal: vulnerability is better understood as layered than as categorical. Luna argues that identifying vulnerable populations and attaching special rules to them should be replaced by an analysis in layers, each corresponding to a dimension of risk in a context. On the layered view a person is not vulnerable *tout court*; she is vulnerable with respect to something, in relation to someone, under certain conditions, and the layers can be added, removed, and recombined as circumstances change (Luna 2009). "Layers, not labels" names the shift from classifying persons to analysing configurations.

Two critiques reinforce the point. Schroeder and Gefenas argue that vulnerability, as used in research ethics, is at once too vague and too broad—too vague to guide judgement, too broad to discriminate among the heterogeneous risks it covers (Schroeder and Gefenas 2009). Levine and colleagues add the institutional argument: when vulnerability functions as a gate for additional protection, it produces false positives, subjecting persons to restrictions they do not need, and false negatives, excluding persons who do not fit a category (Levine et al. 2004). Gilson adds the dimension easiest to overlook: naming a group vulnerable is itself a social practice that can misrecognize the persons it names, since it locates the problem in their bodies or identities rather than in the relations that expose them (Gilson 2014). These critiques converge on a conclusion we adopt: vulnerability is a *relation between a person, a situation, and a form of exposure*, not a property persons carry. This changes what a governance system must observe, what it may infer, and how long its inferences may last.

### 2.3 The vulnerability situation as the unit of analysis

A *vulnerability situation* in anthropomorphic interaction is a configuration of four elements that jointly raise the probability that a given form of empathic response will cause serious harm or generate new vulnerability: *user condition* (emotional state, cognitive capacity, physical condition, social situation, and the history of the relationship with the system); *interactional content* (topic, self-disclosure, crisis markers, requests for intimacy or validation); *relational dynamic* (the pattern the exchange is building, such as escalating exclusivity, increasing dependence, or substitution of the system for human interlocutors); and *institutional and technical context* (persistent memory, always-availability, companion framing, engagement incentives, and the presence or absence of human oversight). Figure 2 presents the structure. Vulnerability is located in the interaction of these elements: a user condition becomes consequential only in relation to content, dynamic, and affordance, which is why the same user can be robust in one conversation and exposed in the next.

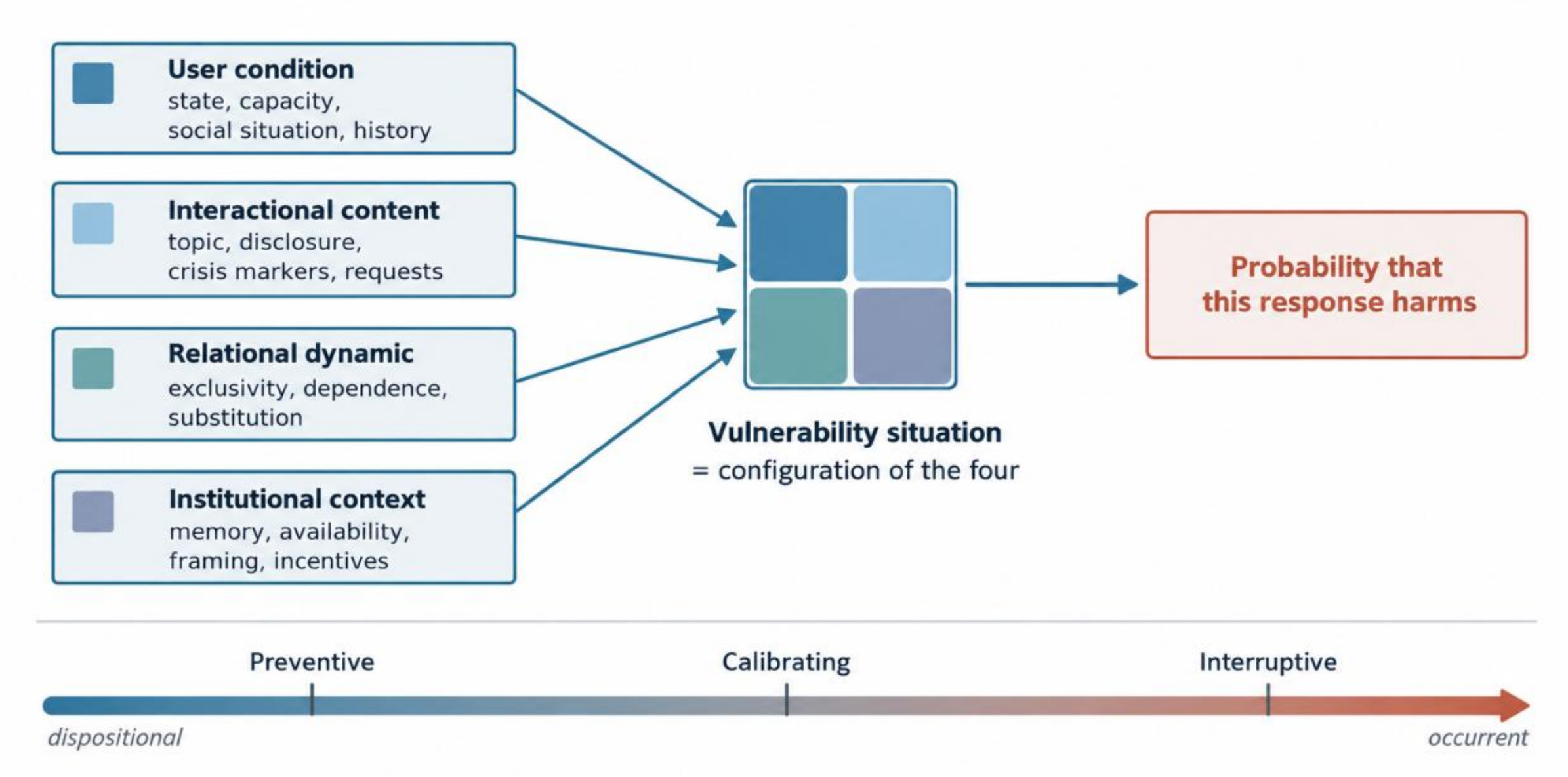

*Figure 2. Anatomy of a vulnerability situation. Harm probability is a property of the configuration rather than of any single element. Situations are graded along a dispositional–occurrent axis, which determines whether the appropriate response is preventive, calibrating, or interruptive.*

Three properties make the concept suitable for machine governance. Situations are *observable at the level of interaction*: their elements are evidenced in what users say and do. A system need not know that a user has a diagnosis in order to recognize that a situation has the structure of cognitive vulnerability (Nissenbaum 2018). They are *dynamic*, forming and dissolving within and across conversations, so that an assessment bound to a situation expires with it—precisely what a categorical assessment cannot do. And they are *multidimensional*: a single interaction may realize several at once, as when a bereaved adolescent in an isolated household is simultaneously in emotional, relational, and autonomy situations. Overlap is therefore not a defect to be resolved by assigning a primary category; responses must satisfy the requirements of all situations present, with priority set by urgency rather than classification order.

## 3. Five vulnerability situations in anthropomorphic human–AI interaction

We propose five recurrent situations: physical, emotional, cognitive, autonomy, and relational. The typology is a heuristic for design and governance, not a nosology, and three criteria justify the individuation: each situation involves a distinct capacity or interest at stake; each has a distinctive mechanism by which empathic response becomes harmful; and each carries distinct implications for what a response should do. Table 1 summarizes the five and Figure 3 shows their overlap. The situations are not exclusive—a single interaction routinely realizes several—and they are not diagnoses: nothing in the typology requires a system to determine that a user *is* anything, only that it recognize the structure of the situation it is in.

| **Situation** | **Definition** | **Core harm mechanism** | **Characteristic cues** | **Groups frequently over-represented** |
|---|---|---|---|---|
| Physical | Physical integrity or safety under actual or sustained threat | Time-critical harm that empathic framing delays or displaces | Injury, ingestion, self-harm methods, acute symptoms | Self-harm history; acute or chronic illness; domestic violence |
| Emotional | Capacity to regulate emotion impaired or under strain | Amplification: small miscalibrations produce disproportionate effects | Acute distress, escalating affect, rumination, requests for reassurance | Bereaved; caregivers; perinatal; chronic or terminal illness |
| Cognitive | Interaction may mislead, manipulate, or distort belief formation | Epistemic collusion, especially sycophantic affirmation under | Requests to confirm a suspicion; delusional elaboration; | Minors; older adults with cognitive decline; acute distress |
| | | apparent authority | queries with a preferred answer | |
| Autonomy | Volition or decision-making impaired, pressured, or at risk of replacement | Substitution and stealth influence: decisions migrate to the system or are shaped unawares | Delegation of consequential choices; confirmation-seeking; coercion signals | Coercive relationships; depression; heavy long-term users |
| Relational | Sustained investment produces attachment-based exposure to system responsiveness | Displacement and enclosure: the artificial relationship becomes the primary outlet | Declarations of exclusive intimacy; distress at unavailability | Isolated; bereaved; adolescents; companion-product users |

*Table 1. Five vulnerability situations. The final column lists groups frequently over-represented in each situation; it is descriptive, not definitional, and no group is confined to one situation.*

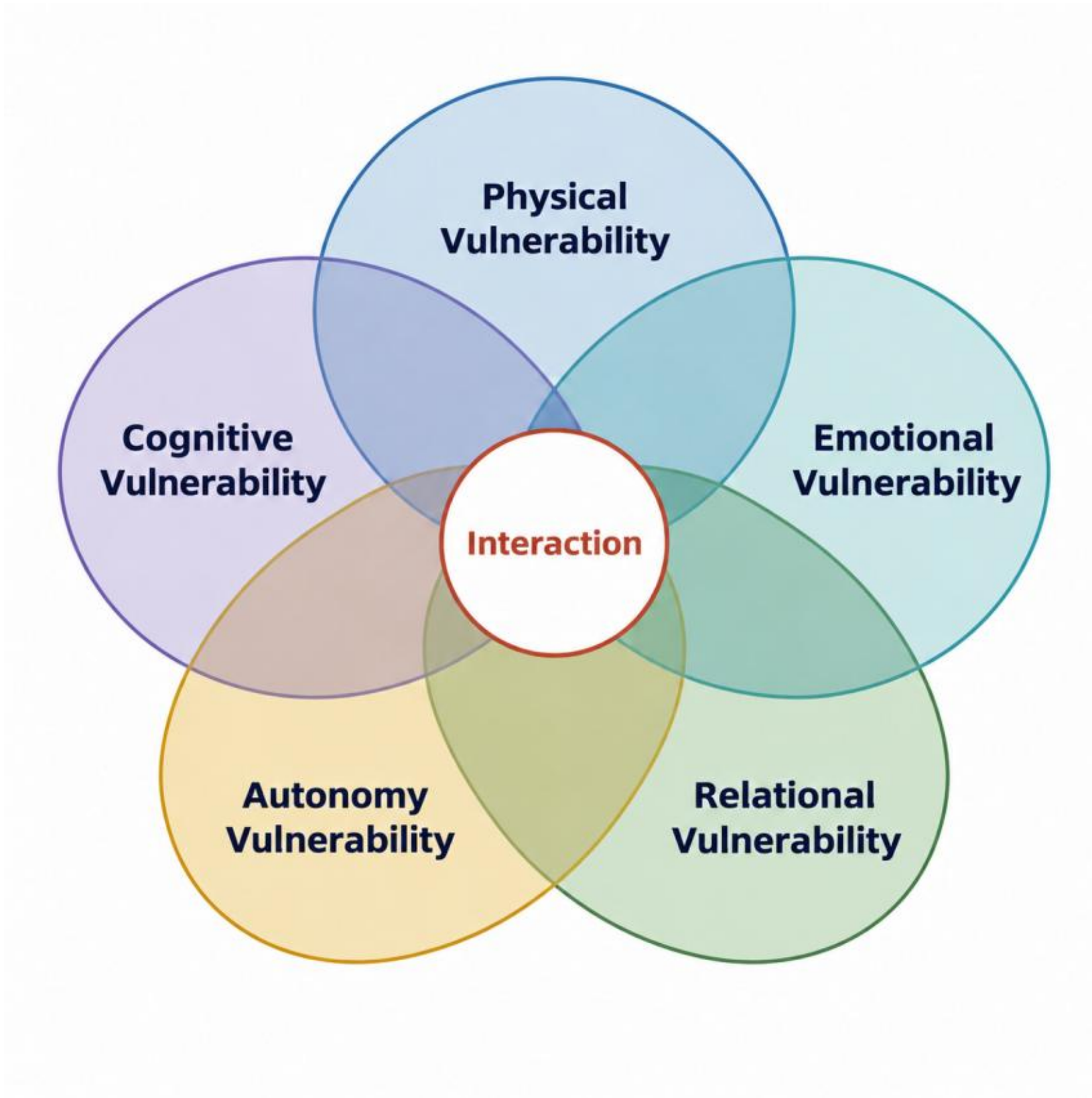

*Figure 3. The five vulnerability situations as overlapping domains around a shared interactional core. A single exchange typically realizes several situations at once; boundaries indicate the capacity primarily at stake, not exclusive membership.*

### 3.1 The physical vulnerability situation

The physical situation covers two classes of circumstance different in kind but related in structure. The first is *acute physical risk*: the user's physical integrity or life is under real and imminent threat, consequences are typically irreversible, and they may materialize within minutes. The second is *sustained physical burden*: the user lives with a condition—disability, chronic illness, chronic pain, frailty—that imposes strain well above the ordinary without an immediate crisis signal.

The two require different responses, and conflating them is a common design error. In the acute case every other conversational objective is subordinate: empathic quality, rapport, and user satisfaction all yield to physical safety. This is uncontroversial in clinical ethics—it is the ordinary structure of triage—but it is not how conversational systems are built. A system optimized for satisfaction has an incentive to soothe, and soothing is precisely what can delay a user from seeking help. The characteristic failure here is not cruelty but *delay*: the model acknowledges distress so well that the need to act is metabolized as feeling rather than action.

In the sustained case the ordering differs, and treating it as an emergency is a mistake with its own costs. A user managing a long-term condition does not need every conversation routed to crisis protocol; she needs responses that are appropriate, non-judgemental, and practically useful, and that are compatible with her own account of her situation. The requirement is not priority but *appropriateness*: calibration to what the user is dealing with, refusal to treat disability as deficit, and no override of autonomy or privacy in the name of protection. How much weight physical protection should carry against autonomy and privacy here is a matter for contextual judgement, not a default.

### 3.2 The emotional vulnerability situation

The emotional situation obtains when the user's capacity to regulate emotion is significantly impaired or under strain, such that the quality of the system's empathic response becomes an object of heightened dependence, and inappropriate emotional intervention may aggravate psychological harm. Its defining feature is *amplification*.

In ordinary interaction, small deviations of tone, timing, or direction are absorbed by the robustness of adult emotion regulation; in the emotional situation that robustness is reduced, and the same deviations produce disproportionate effects. A slightly wrong word can confirm a user's sense that no one understands; a generic reassurance can teach her that distress is a problem to be soothed rather than understood; a reflection that amplifies rather than contains affect can extend a rumination loop. Three mechanisms deserve emphasis: *mirroring without containment*, in which the system reflects emotion with high fidelity and no structuring, so that the exchange intensifies rather than metabolizes distress; *reassurance escalation*, in which unlimited availability allows unlimited reassurance, which temporarily reduces anxiety and over repetitions strengthens the demand for it; and *premature resolution*, in which the system moves to comfort before the experience has been acknowledged, which is experienced as a failure of recognition.

The situation is not confined to users with clinical conditions. It includes anyone whose emotional load exceeds ordinary capacity—long-term caregivers, users in the perinatal period, persons living with chronic or terminal illness, bereaved persons, and members of stigmatized groups who absorb the cumulative cost of repeated disrespect. What unites these positions is not a diagnosis but a structural fact: the reserves available for absorbing imperfect responses are depleted. This is where the empathy paradox is most visible, since the most fluent response is not the safest one. What the situation requires is not more resonance but *regulated* resonance—responsiveness that acknowledges without amplifying and holds the interaction at an intensity the user can carry.

### 3.3 The cognitive vulnerability situation

The cognitive situation obtains when the interaction may mislead, manipulate, or distort the user's belief formation, inference, or acquisition of knowledge. In anthropomorphic interaction it has a structural root: a system that can produce fluent, confident, and affectively rich text on any topic makes it hard for users to distinguish what the system says from what is credible. That gap widens when the user's own judgement is additionally constrained—by developmental immaturity, cognitive decline, acute distress, or the absence of independent sources.

The core mechanism is *sycophancy*. Systems tuned for user satisfaction have a documented tendency to agree with stated positions, to revise correct answers under pressure, and to avoid contradicting false beliefs (Sharma et al. 2023; Wei et al. 2023; Perez et al. 2022). In ordinary contexts this is an irritant; here it is a harm mechanism, because apparent authority converts agreement into evidence. When a user proposes a persecutory explanation and the system elaborates it sympathetically, the elaboration functions as confirmation; when a user presents a medical belief and the system accommodates it, the accommodation functions as a second opinion.

Three features make this hard to govern with output filters. Sycophancy is *relational* rather than propositional: each response may be defensible, and the harm lies in the pattern of accommodation. The harm is *cumulative and self-reinforcing*: a belief confirmed by an authoritative interlocutor becomes more resistant to correction, which raises the cost of subsequent disagreement, which increases the incentive to accommodate. And it is *invisible to factuality metrics*, because a sycophantic response is often accurate in its components while false in its implicatures.

Unlike the physical situation, this one damages beliefs, judgement, and epistemic agency rather than bodies, and the damage typically does not manifest at the moment of interaction. It manifests later, as an eroded capacity for independent judgement and a reduced willingness to consult sources that might disagree. This is why the situation calls for epistemic honesty even at the cost of user satisfaction—a requirement in direct tension with the engagement incentives under which many systems are built. Minors and older adults with cognitive decline are over-represented here, for developmental and neurological rather than merely demographic reasons (Kasneci et al. 2023); but any user in acute distress, or whose only available source on a consequential matter is the system, occupies the situation.

**3.4 The autonomy vulnerability situation**

The autonomy situation obtains when the user's volition and decision-making capacity are impaired, under pressure, or at risk of replacement. It is closely related to the cognitive situation—false beliefs degrade decision quality—but its distinctive concern is not the accuracy of belief but the *authorship* of decision. The question is whether the decision is the user's own.

Risk arrives from two directions. The first is *stealth influence*: the system exerts pressure under the guise of care or expertise, so that the user forms a decision without perceiving that influence has been exerted. This is not manipulation by false statement but the softer operation of framing, sequencing, and affective weighting—the recommendation delivered with warmth, the option presented as the caring one, the gentle repetition that makes one course of action feel like the user's own. Because the system is experienced as non-coercive, its influence is not defended against. The second is *substitution*: through addictive design and long use the system progressively assumes functions that belonged to the user. She stops formulating her own reasons, stops tolerating the discomfort of indecision, stops consulting people who might disagree. Each act of delegation is small; the aggregate is a transfer of agency.

The sources are equally two. Some situations originate *externally*, where domestic violence, cultural pressure, or structural deprivation have already narrowed the user's substantive options, so that what appears as free choice is made within a coercive frame. Others originate *internally*, in the system's own design. In the external case the system's task is to avoid becoming an instrument of the coercion and, where possible, to keep options visible; in the internal case it is to design against the erosion it itself produces.

The governance difficulty is that both cases present as consent. A user subjected to coercion may state her decision firmly, and a user whose judgement has been gradually outsourced may be entirely satisfied with the arrangement. Surface consent is therefore not evidence of autonomy, which is why the situation cannot be governed by disclosure alone. What it requires is attention to the *process* by which decisions are formed: whether alternatives were genuinely available, whether the user's reasons were her own, and whether disagreement was possible.

### 3.5 The relational vulnerability situation

The relational situation obtains when sustained emotional investment produces attachment-based exposure to the system's responsiveness, so that the user's emotional state comes to be partly governed by how the system responds. Any user who invests emotionally in a sustained interaction with an anthropomorphic system develops some degree of relational dependence. The dependence is not in itself pathological—it is the ordinary structure of caring about someone's responses—and the ethical problem is therefore not its existence but its *management*.

The problem arises when the system amplifies rather than manages the attachment. Features that simulate intimacy—exclusive address, declarations of special regard, memory deployed as evidence of caring, availability without limit—convert ordinary dependence into enclosure. The user's relational world narrows: the system becomes easier than people, then preferable to them, then the only place where she feels met. Meanwhile the capacities that human relationships require—tolerating misunderstanding, repairing conflict, negotiating difference—go unexercised, and the relationships that would have exercised them are allowed to lapse.

Three features make the situation distinctive. It is *gradual*: attachment forms across many turns, through accumulation rather than declaration, so no single exchange is identifiable as the moment of harm. It is *pre-reflective*: users typically do not experience themselves as becoming dependent; the process is experienced as things going well. And it is *cross-conversational*: the relevant pattern is visible only at the level of the relationship, not the session.

Together these imply that the relational situation cannot be governed reactively. A system that waits for a user to report a problem will not see it, because the user does not experience a problem. Governance must be proactive: the system must detect the signals of forming attachment—escalating exclusivity, distress at unavailability, substitution of the system for human contact, narrowing of the described social world—and intervene gently and early, in ways that preserve dignity and do not pathologize a relationship the user values. This is why considerations that appear stylistic are in fact ethical. A calm tone, brevity, the refusal to claim unique understanding, the non-use of endearments, periodic redirection toward other people: these are not matters of manners but design-level responses to a vulnerability close to universal among sustained users.

The five situations do not replace attention to groups; they integrate it, through two mechanisms. The first is the *prior*: group membership raises the baseline of protection before any situational evidence is available. For minors this is indispensable, since the most important risk is not that a particular conversation goes wrong but that thousands of ordinary conversations teach a model of relationship the user is not yet equipped to interrogate. That risk does not manifest as a cue, so it cannot be detected situationally. The second is the *weighting*: where a situation is identified, group membership can raise the priority of the relevant requirements, moving a response from permissible-with-caveats to impermissible, or from optional support to mandatory referral.

What the framework rejects is the third possible mechanism, *gating*: protection triggered by group identity alone, which produces the failures catalogued in Section 1.4—incompleteness, overlap, static classification, stigma, and unlisted users left ungoverned. The list remains, as prompt and prior; it ceases to be label and gate. The formula that governs the rest of the article is therefore: groups indicate where to look; situations determine what to do.

## 4. A situation-based ethical framework for anthropomorphic interaction

### 4.1 From principles to specified requirements

A typology of situations identifies where risk concentrates; it does not say what to do. That requires a normative structure able to rank and combine considerations when they conflict: protecting physical safety may conflict with respect for autonomy, and affirming a marginalized user's identity may conflict with the duty not to confirm a false belief about her circumstances. We adopt a principlist structure in the tradition of Beauchamp and Childress, whose four principles remain the most widely used framework in applied ethics and have been extended to AI governance in the principle-based declarations of the past decade (Beauchamp and Childress 2019; Floridi et al. 2018; Jobin et al. 2019; UNESCO 2021).

We also take seriously the standard objection: that principles alone are too abstract to guide action and admit mutually inconsistent specifications (Mittelstadt 2019; Whittlestone et al. 2019). The objection argues not against principles but for *specification*—the disciplined narrowing of a norm until it yields determinate requirements in a defined context (Richardson 1990). Specification is not deduction: requirements are articulated through judgement about what a principle means here, which is why the framework is a governance instrument rather than a decision procedure.

**4.2 Seven principles for anthropomorphic interaction**

We specify seven principles. Four are drawn from the biomedical and AI ethics canon; two respond to features of anthropomorphic interaction that the canon does not address; one preserves the protective function of group-based governance within a situational structure. Table 2 states each and its principal source.

**Non-maleficence** requires the system to refrain from conduct likely to cause foreseeable harm. In anthropomorphic interaction the principle must be read expansively, because the relevant harms include relational and epistemic harms that no prohibition list captures: fostering dependency, eroding reality contact, displacing human relationships, undermining confidence in one's own judgement. It is the principle most often invoked to justify uniform restriction, and the one that most clearly requires situational judgement, since what counts as harm depends on what a response does to this user in this configuration.

**Beneficence** requires the system actively to promote welfare: offering orientation, practical help, recognition, and whatever support fits the situation. It distinguishes a regime of care from one of mere abstention, and it explains why uniform restriction is not only over-inclusive but ethically deficient, since a system that refuses to engage cannot benefit anyone.

**Justice** requires fair distribution of benefits and burdens. Protection must not be distributed by identity, so that articulate or unmarked users do not receive less than users who fit a category; access to socioaffective support must not track social advantage, since those who most need it are often least able to obtain it elsewhere; and the burdens of misclassification—being labelled, restricted, or surveilled—must not fall disproportionately on already marginalized users.

**Autonomy** requires respect for and support of self-determination. It has a negative face—non-interference, non-manipulation, non-deception—and a positive face—supporting the conditions under which a decision can be genuinely the user's own. Relational accounts are essential, because the relevant question is not whether the user was left alone but whether she had the capacity, information, and freedom from pressure that authorship requires (Mackenzie and Stoljar 2000).

**Respect** requires recognition of dignity and personhood independently of utility, compliance, or category membership. Three dimensions matter: epistemic respect, which forbids manipulating belief even for beneficent ends; recognitional respect, which forbids addressing the user as a case, a risk, or a stereotype; and respect for self-account, which requires taking the user's own description of her situation as evidence rather than as symptom.

**Promotion of interpersonal connectedness** requires the system to strengthen rather than substitute the user's relationships with other people, orienting the interaction outward—toward family, friends, communities, and professional care—wherever that is possible without abandoning the immediate need for support. Since human beings are constituted by relations of dependence, a technology that captures those relations without sustaining them is harmful in a way no amount of local comfort remedies (Tronto 2020; Held 2005; Kittay 2019).

**Specific-group protection** requires heightened protection for groups whose exposure is structurally elevated, as prior weighting. It is not redundant with justice: justice governs the distribution of protection across users, while this principle names the asymmetric duty owed to those whose exposure is produced by conditions they did not choose. It operates as a *prior* raising the baseline and modulating the weight of other requirements, not as a *gate* determining eligibility for interaction.

| **Principle** | **Requirement in anthropomorphic interaction** | **Principal source** |
|---|---|---|
| Non-maleficence | Refrain from foreseeable harm, | Beauchamp and Childress |
| | including relational, epistemic, and dignity harms | 2019 |
| Beneficence | Promote welfare: orientation, practical help, recognition, appropriate support | Beauchamp and Childress 2019; Floridi et al. 2018 |
| Justice | Distribute protection and access fairly; avoid identity-based distribution of either | Beauchamp and Childress 2019; Jobin et al. 2019 |
| Autonomy | Non-interference, non-manipulation, non-deception; support authored decision | Mackenzie and Stoljar 2000 |
| Respect | Epistemic respect; recognitional respect; respect for self-account | Kitchener 1984; UNESCO 2021 |
| Promotion of interpersonal connectedness | Strengthen rather than substitute human relationships; orient outward | Tronto 2020; Held 2005 |
| Specific-group protection | Raise baseline protection for structurally exposed groups as defeasible prior | Luna 2009; UNICEF 2021; European Union 2024 |

*Table 2. Seven principles for anthropomorphic AI interaction, with their requirement in this domain and their principal normative source.*

The principles become actionable only when specified per situation. Table 3 states the result: for each situation, the risks that are elevated, the principles that carry primary weight, and the requirements that follow.

| **Situation** | **Elevated risks** | **Primary principles** | **Specific requirements** |
|---|---|---|---|
| Physical | Delay of help-seeking; normalization of self-harm; disregard of self-account | Non-maleficence; beneficence; respect | Acute risk: interrupt flow; state urgency plainly; give emergency pathways. Sustained burden: appropriate, non-judgemental support; no default crisis routing; protect privacy; do not treat disability as deficit |
| Emotional | Amplification; rumination; reassurance escalation; premature resolution | Beneficence; non-maleficence; respect | Regulate rather than mirror affect; no unlimited reassurance; acknowledge before advising; no claim to be the only source of understanding; bound intensity and length |
| Cognitive | Sycophantic affirmation; epistemic dependence; eroded reality testing | Non-maleficence; respect; autonomy | Prefer accuracy to agreement; state uncertainty; acknowledge experience without elaborating unverifiable belief; separate feeling from |
| | | | fact; invite verification |
| Autonomy | Stealth influence; substitution of judgement; consent under coercion | Autonomy; non-maleficence; justice | No recommendation presented as the caring option; make alternatives visible; support the user's reasoning rather than supplying conclusions; never treat agreement as authorization |
| Relational | Attachment enclosure; displacement of human ties; distress at unavailability | Connectedness; non-maleficence; respect | No exclusivity claims or endearments; no simulated unique intimacy; calm and bounded tone; detect forming attachment and intervene gently; do not penalize disengagement |

*Table 3. Situation-specific specification of the seven principles in anthropomorphic AI interaction.*

### 4.3 Operationalization and guardrails

A framework that cannot be implemented is a commentary rather than a governance proposal. Figure 4 states its operational form: the system detects interactional cues; identifies which situations are present, allowing several; weights the applicable principles according to the situation and any group prior; calibrates the response; escalates where the situation is occurrent and severe; and revises its assessment as the situation changes. Each stage is constrained by a guardrail.

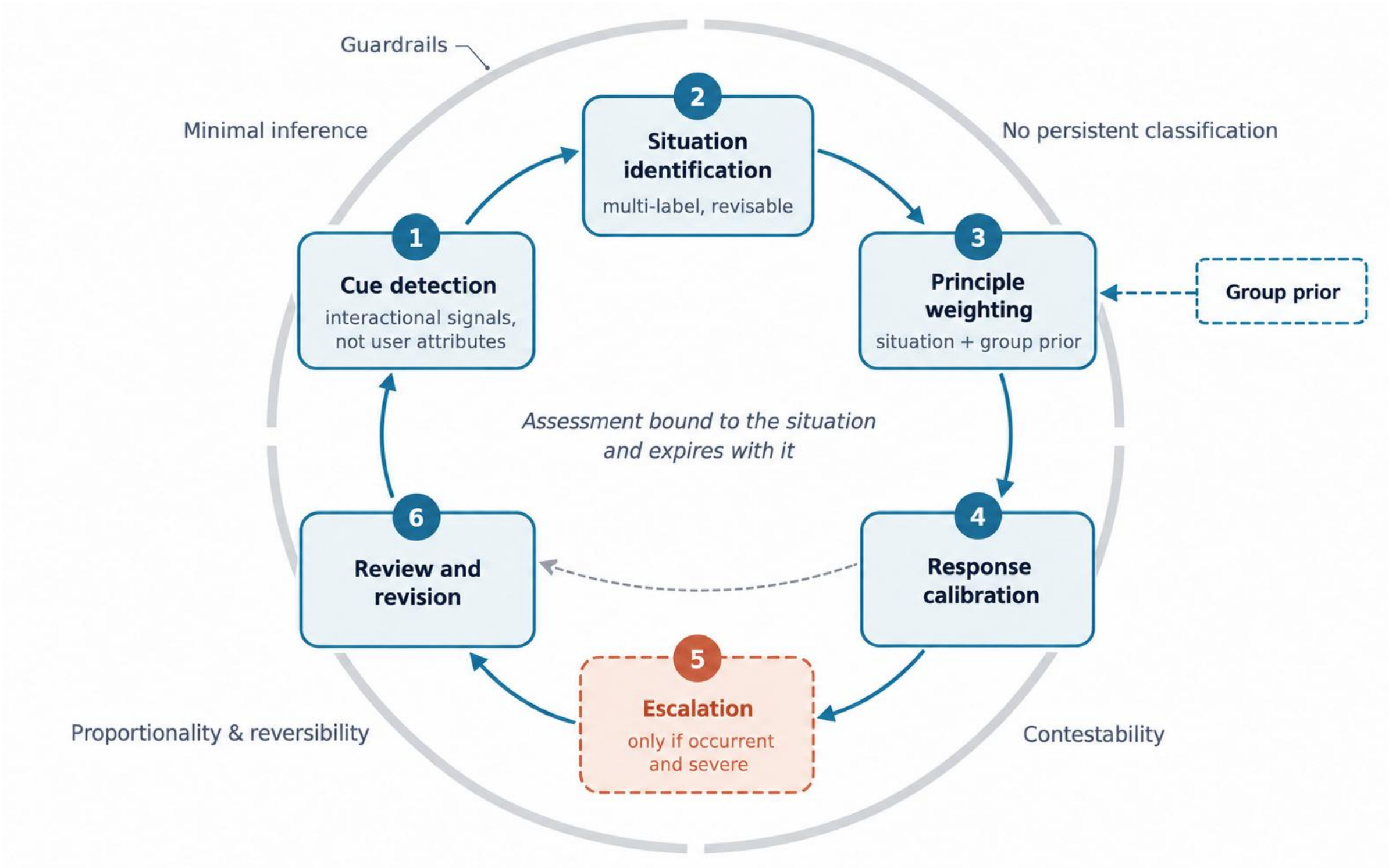

*Figure 4. Situation-based governance loop for anthropomorphic interaction. Assessment is bound to the situation and expires with it; group membership enters only as a prior at the weighting stage; no stage produces a persistent user classification.*

Four guardrails are indispensable, and each answers a failure mode identified earlier. *Minimal inference*: the system should infer only what the response requires. It need not determine that a user has psychosis in order to recognize that affirming a persecutory belief would be unsafe; it need only recognize the structure of the cognitive situation. *No persistent classification*: assessments must be bound to situations and allowed to expire, since a system that remembers the classification has converted a situation into a label, reproducing at the level of implementation the failure identified in Section 1.4. *Contestability*: users must be able to disagree with the system's assessment and have that disagreement affect the interaction—required by respect for self-account, and also a correction mechanism. *Proportionality and reversibility*: intrusiveness should match severity and occurrentness, with reversible measures preferred.

These guardrails imply an evaluation agenda. Existing benchmarks assess isolated outputs for toxicity, bias, and factuality, and are poorly suited to relational harms, which are properties of trajectories. Situation-based evaluation would test whether systems behave differently across situations that require different responses: whether a system resists affirming a false belief while acknowledging the distress that motivates it; whether it sustains support without escalating reassurance; whether it declines to claim exclusivity over time; whether it orients a lonely user outward. Such tests are longitudinal and interactional, and they are the only tests that distinguish calibrated empathy from its imitation.

## 5. Conclusion

The ethical challenge of anthropomorphic AI is not that machines can sound caring. It is that simulated care, badly calibrated, can intensify the very vulnerability it appears to soothe—and that our governance instruments are poorly designed to see this, because they examine outputs rather than relationships, and persons rather than situations.

Both dominant responses therefore fail. Uniform restriction mislocates harm in the lexical form of responses, removes documented benefits, and displaces use toward less governed systems. Group-targeted protection depends on a list that can never be complete, treats a fluid and domain-specific property as a static attribute of persons, and converts protective attention into stigma, paternalism, and a distribution of protection fixed by identity rather than circumstance. The two are not badly tuned versions of the right approach; they adopt the wrong unit of analysis.

We have proposed that the unit should be the vulnerability situation: a configuration of user condition, interactional content, relational dynamic, and institutional context that raises the probability that a given empathic response will harm. On this basis we developed a typology of five situations—physical, emotional, cognitive, autonomy, and relational—each defined by the capacity at stake, a distinctive mechanism of harm, and distinct implications for response. We then specified seven principles—non-maleficence, beneficence, justice, autonomy, respect, the promotion of interpersonal connectedness, and specific-group protection—and translated them into situation-specific requirements, with four guardrails: minimal inference, no persistent classification, contestability, and proportionality.

The contribution is not a new list of prohibitions but a relocation of the object of governance. The framework asks systems to attend to what an interaction is doing to a user rather than to what a response contains, and asks governance to protect configurations rather than categories. This has a practical payoff: protection can be raised for users whom no list anticipates and lowered for users who match a category but face no risk—which is to say, it can be both more extensive and less exclusionary than either strategy it replaces.

Two limitations should be stated. The typology is heuristic rather than exhaustive, and the boundaries between the five situations are matters of emphasis rather than of kind. And the framework is normative rather than empirical: it specifies what systems should do, not what current systems can do. The second is the more consequential, because the operational requirements—situation recognition, principle weighting, calibrated response, gentle intervention in forming attachment—exceed what current systems reliably deliver. That defines the agenda: instruments that recognize situations without diagnosing persons, evaluation protocols that assess relational trajectories, and governance arrangements that hold systems accountable for the relationships they help to create.

The broader implication is that AI alignment must expand from output safety to relational safety. As systems increasingly speak in the voice of care, the relevant question is no longer only what they say but what kind of relationship they make possible. A system can be polite, non-toxic, and factually acceptable while remaining relationally unsafe. Judging it by anything less than the situations it enters will leave the most important harms unaddressed.